\documentclass[twocolumn,showpacs,preprintnumbers,amsmath,amssymb]{revtex4}

\usepackage{graphicx}
\usepackage{dcolumn}
\usepackage{bm}

\begin{document}
\title{Empirical Formula of the Absolute Value of Electrical
  Conductivity for Elemental Metals and Its Interpretation By Fluctuation Dissipation Theorem}

\author{Tadashi HIRAYAMA}
\affiliation{National Astronomical Observatory of Japan, Mitaka, Tokyo
  181-8588, Japan}
\date{\today}

\begin{abstract}
We propose an empirical formula of the absolute values of electrical conductivity $\sigma$
for pure elemental metals such as Na, Cu, or Fe
at room temperature ranges. Assuming the relaxation time of $\tau_{0}=\hbar/k_{\rm B}T$ for all metals,
we propose $\sigma=e^{2}n_{\rm atom}\tau_{0}/(mG)$ ($n_{\rm atom}$
=number density of atoms in each metal, but not that of electrons $n$, $m$=true electron mass).
If we adopt that a single free parameter $G$ is the sum of outer electron numbers in electron
configuration such as $G=\!1+2\!=\!3$ 
 for In$^{49}$(5s$^{2}$4p$^{1}$), the `absolute values' of
$\sigma$ and the thermal conductivity $\lambda$ agree with experiments
within $\sim 20{\%}$ for the most of metals, including semimetals (Bi, Sb, and As), and
also, only for $\lambda$, Si and Ge.

We find that the above results are only compatible with
$n=n_{\rm atom}$ ($Z=1$) for all metals as far as $\sigma$ is concerned,
 and also compatible with $G/\tau_{0}=\Sigma_{i=1}^{G}(1/\tau_{i})$.
Here $\tau_{i}=\tau_{0}$ is assumed for each $i$ band; 
e.g. each of two 5s and one 1p bands in Indium.
Using the theoretical state density,
the electron specific heat ratio $C_{\rm Ve-obs.}/C_{\rm Ve-theory}$
leads to $m^{\ast}/m\simeq 1$ for 24 metals,
supporting use of $m (m^{\ast}$=effective `thermal' electron mass). 
We derive $\tau_{0}=\hbar/k_{\rm B}T$ by (A)$\sim$(C) methods. (A) The
Bardeen's (1937) $\tau$ is reduced to $\tau_{0}$ by adopting an
empirically ascertained relation of the acoustic and Fermi energy, 
though requiring the deformation potential equal to the Fermi energy.
(B) Assuming that $1/\tau=n_{\rm eff}SV_{\rm F} (V_{\rm F}=$Fermi
velocity, $S$=collision area), and $n_{\rm eff}$ is the effective
number density of colliding electrons due to the Fermi-distribution saturation, we can reproduce $\tau_{0}$.
(C) To obtain $\tau_{0}$, the fluctuation-dissipation theorem is found to require\,
$\int\frac{1}{2} m<v(0)v(t)>dt=<\Delta E><\Delta t> =\hbar/2$.
Here $v(t)$ is the fluctuating electron velocity around the Fermi velocity. 
This $\hbar/2$ is the Heisenberg's minimum uncertainty value,
related to the minimum wave function, which in turn is realized
by free electron motions suggested by very large mean free paths in
metals. The version (C) needs no parameters and seems a most
reliable.
\end{abstract}

\maketitle

\begin{center}
\bf{ 1.  Introduction} 
\end{center}
\newcommand{\mfp}{$L_{\rm mfp}$}
\newcommand{\kB}{$k_{\rm B}$}
\newcommand{\natom}{$n_{\rm atom}$}
\newcommand{\VF}{$V_{\rm F}$}
\newcommand{\EF}{$E_{\rm F}$}
\newcommand{\kF}{$k_{\rm F}$}
\newcommand{\hbarkbt}{$\hbar/k_{\rm B}T$}
\newcommand{\Schrodinger}{$\rm{Schr\ddot{o}dinger}$}

In this paper we treat only elemental metals 
 such as Na, Cu, Fe, etc. without impurity and in the room temperature range.
 We study metals of normal geometrical size at standard pressure. 
Conventional formulae \cite{ashcroft,ibach,kittel} for the electrical
 conductivity $\sigma$ and thermal conductivity $\lambda$ in the free
 electron model are $\sigma=(\tau n/{m^\ast})e^{2}$ and
 $\lambda=(\tau n/m^{\ast})\pi^{2}k_{\rm B}^{2}T/3$, respectively. 
 Here $\tau$ is the relaxation time of electrons 
 at the Fermi energy $E_{\rm F}$, $m^{\ast}$ is the
 effective electron mass and $n$ is the electron number density of metals,
 satisfying $n=Zn_{\rm atom}$, where $n_{\rm atom}$=metal density/atom weight [$\rm m^{-3}$].
 In order to derive `the absolute values' of
 $\sigma$ and $\lambda$ one needs to know $\tau$, $Z$ and $m^{\ast}$,
 all of which are poorly known for many metals, especially $\tau$, if
 not from the observations. 

For example, we note that $\tau\sim \hbar/k_{\rm B}T$  was 
claimed for $T\gg \Theta$ \cite{lifshitz,abrikosov}  
and Abrikosov \cite{abrikosov} extends to use it also for $T\approx \Theta$  as in eq.(4.18) (see foot note therein). 
Here $\Theta$ is the Debye temperature. The accuracy of the `tilde' signs they used is, however, not clear, 
but also they did not treat multivalent metals.
Pippard \cite{pippard} holds a high opinion of the 1937-Bardeen \cite{bardeen} calculation for monovalent metals 
on the absolute $\sigma$ value of Na and K, 
while Ziman \cite{ziman} expressed that it is not very accurate, probably because of a factor of two to three difference 
between experiments of $\sigma_{\rm obs}$
and the Bardeen theory for Rb, Cs, Cu, Ag, and Au, where $\sigma \propto$ ($E_{\rm F}/C)^{2}$. Here $C$ is the deformation potential.
A concise derivation of the relaxation time $\tau\equiv 1/W$ is in Kittel \cite{kittel} (Appendix J)
which still needs values of  $C, m^{*}/m$ and the sound speed.
Here again no statements are made on multivalent metals. 
Certainly, there have been attempts
to relate $\sigma$ to electronic configuration early in 1950's e.g. by Gerritsen\cite{gerritsen}, but combination of parameters are unlike
the present ones, which are $G$ and $n_{\rm atom}$ as shown below; namely these attempts were effectively unsuccessful.

This paper is organized as follows. Section 2 presents our formulae for
$\sigma$ and $\lambda$. Section 3 compares the experimental values with our formulae,
by assigning the single parameter $G$ of the sum of outer eletron bands in each atom.
Sections 4-8 inspect if the assumptions made in Sect.2  are feasible.
Section 4 finds $n=n_{\rm atom}$, namely $Z=1$, for all elemental metals (not for alloys), though $n$-values were
not needed in Sects.2 and 3, and discusses the role of $G$ in the
macroscopic momentum equation.
This $Z=1$ is applicable only for $\sigma$ and $\lambda$.
Section 5 finds that $m^{\ast}/m\simeq 1$ is also consistent with the electron specific heat expected from the theoretical
Fermi energy.
Section 6 finds that the $\tau$-value given by Bardeen \cite{bardeen} is 
reduced to $\tau_{0}\equiv$\hbarkbt \, if an empirically ascertained relation
of $3MC_{\rm s}^{2}\approx$ \EF is employed, which resembles the Bohm-Staver relation ($M$=ion mass and $C_{\rm s}$=total sound speed).
Section 7 derives an approximate value of $\tau_{0}$ from $\tau^{-1}=n_{\rm eff}S$\VF \,`without explicitly introducing ion thermal vibrations'. Here
$n_{\rm eff}\propto n \frac{k_{\rm B}T}{E_{\rm F}}$ is the effective number density of electrons in the Fermi distribution,
$S$ is the collision cross-section and \VF\, is the Fermi velocity.  
Section 8 adopts that $\sigma=ne^{2}\int_{0}^{\infty}<v(0)v(t)>dt/k_{\rm B}T$ from the classical fluctuation dissipation theorem of Kubo\cite{kubo} 
is equal to $\sigma$ from the Drude form ($v(t)$ is the fluctuating electron velocity), 
and then finds that in order to reproduce $\tau_{0}=\hbar/k_{\rm B}T$, 
$\frac{1}{2}m<v^{2}>\Delta t=\hbar/2$ should hold, claimed earlier 
by the present author\cite{hirayama}. Since electrons in metals are considered
to behave as free electrons as judged from the large mean free paths,
their wave functions will take the minimum uncertainty  Gaussian form
so that the minimum uncertainty relation will hold, consistent with the requirement from the observations. 
Section 9 is the discussion and Sect.10 is the summary. 
Appendix gives the standard derivation of the Drude
formula from the Boltzmann equation with some discussions.\\

\begin{center}
\bf{ 2. Formulae We Propose} 
\end{center}

 In this paper we give these absolute
 values which are in good accord with the observations by adopting
 assumptions below.
We assume for `all' elemental metals
\begin{equation}
\tau_{0}=\hbar/k_{\rm B}T,\label{Eq.1}
\end{equation}
and introduce a non-dimensional parameter $G$ (integer) in place of the conventional
$m^{\ast}/(mZ)$ appearing in the Drude formula. We then propose, using $n_{\rm atom}$ instead of the electron density $n$,
\begin{equation} 
\left(\begin{array}{c}\sigma\\\lambda \end{array}\right)
=\frac{e^{2}n_{\rm atom}\tau_{0}}{m}\frac{1}{G}\left(\begin{array}{c}1\\ \pi^{2}k_{\rm B}^{2}T/(3e^{2}) \end{array}\right).
\label{Eq.2}
\end{equation} 
Here $\sigma$ is in $\rm\Omega^{-1}m^{-1}$, $\lambda$ is in $\rm Wm^{-1}K^{-1}$, $\hbar=h/2\pi (h$ is the Planck 
constant), $k_{\rm B}$ is the Boltzmann constant, and $e$ is the elementary charge (positive value).

These two equations, Eqs.(1) and (2) `combined', do not seem to have been proposed in the past. 
Note that Eq.(\ref{Eq.2}) gives $\sigma \propto 1/T$ due to $\tau_{0}\propto 1/T$ and temperature
`independent'-$\lambda$ both being consistent with observations in the
room temperature range (`independent' means as compared to $\sigma
\propto 1/T$ variation). Since $n_{\rm atom}$ for each metal (from density and atomic weight) and $T$ can be given, the
only non-dimensional parameter is $G$. Even if $\tau$ is different from Eq.(\ref{Eq.1}), unknown departure factor from it  
can be included in $G$ (e.g. $m^{\ast}/m$ or deformed potential in a non-dimensional constant).
Thus assuming $\tau_{0}=$ \hbarkbt,
we first empirically determine the parameter $G_{\rm obs}$ 
using observed $\sigma_{\rm obs}$ for each metal. Then we assign
$G$(guessed)-values to be the sum of the outer electron numbers in
electron configuration which are `close' to $G_{\rm obs}$.\\ 

\begin{center}
\bf{ 3. Comparison with Experiments} 
\end{center}

Figure \!1(a) presents $G_{\rm obs}$ plotted against `group' number for each
`period' in the periodic table. Here $G_{\rm obs}$ is defined as
\begin{equation}
G_{\rm obs}\equiv \frac{\sigma_{1}}{\sigma_{\rm obs}}\;\,\,\mathrm{and}\quad 
 \sigma_{1} \equiv \sigma(G=1)=\frac{e^{2}n_{\rm atom}}{m}\frac{\hbar}{k_{\rm B}T}. \label{Eq.3}
\end{equation}
Then $G_{\rm obs}$ can be given for each metal from $\sigma_{\rm obs}, n_{\rm atom}$ and
temperature $T$ used in the observations. Observed values
($\sigma_{\rm obs}$ and $\lambda_{\rm obs})$ for 48 metals are taken from
Kittel\cite{kittel}, adding $1/\sigma$(As$^{33}$)=333n$\Omega$m and $\lambda$(Ca)=201Wm$^{-1}$K$^{-1}$ from the table of Phys. Soc. Japan
\cite{psj} (PSJ-table). We adopt $T_{\rm obs}$=295K from the
Kittel's tabulation for $\sigma_{\rm obs}$. This gives $\tau_{0}=2.59\times 10^{-14}$s from Eq.(\ref{Eq.1}),
which is very close to $\tau_{\rm obs}$ from the observed $\sigma_{\rm obs}$ such as 
$\tau_{\rm obs}$(Na)=2.9$\times 10^{-14}$s and $\tau_{\rm obs}$(Cu)=2.5$\times10^{-14}$s,
using Eq.(2) for $G=1$.  

First, we find similar trends of $G_{\rm obs}$ among periods of 4
(K$^{19}-$Ga$^{31}$), 5 (Rb$^{37}-$Sb$^{51}$), and 6
(Cs$^{55}-$Bi$^{83}$, inclusive of La$^{57}$). This suggests that
electron configuration, which is the basis of the periodic table, may be
responsible (see e.g. early attempts in Figs. 12-13 by Gerritsen \cite{gerritsen}).
Secondly, $G_{\rm obs} \approx 1$ is found for Na$^{11}$,
K$^{19}$, Rb$^{37}$, Cs$^{55}$, and noble metals of Cu$^{29}$, Ag$^{47}$,
and Au$^{79}$. This indicates that Eq.(\ref{Eq.2}) for the electrical
conductivity $\sigma$ with $G=1$ agrees with the observations without
further parameters (Fig. 1(b)-upper). 
These seven elements in free atomic form have each $3s^{1}-6s^{1}$ outermost electron.
Though one might say that $Z=1$ and $m^{\ast}=m$ hold as 
expected, $\tau_{0}$ should be specified as we propose. 

Third, many metals appear concentrated in $G_{\rm obs}=1$, 3, 5 and $10- 
13$, which suggests discreteness of $1/\sigma_{\rm obs}$ if expressed in
unit of $1/\sigma_{1}$, namely the true $G$ may well be quantized!

\begin{figure}[t]
\begin{center}
\includegraphics[width=7.5cm]{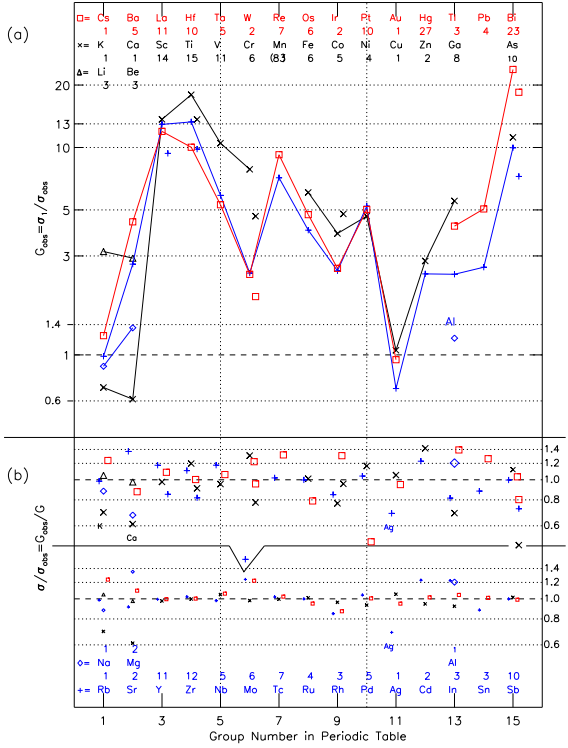}
\end{center}
\caption{(color online) (a) $G_{\rm obs} \equiv \sigma_{1}/ \sigma_{\rm obs}$ and (b) $G_{\rm obs}/G
=\sigma/\sigma_{\rm obs}$ against group number for each period, both in
the same logarithmic scales. Thick marks in (a) and (b)-upper are
 $\lambda_{1} / \lambda_{\rm obs}$ values for eight metals with
 $\vert C_{\rm WF}-1 \vert > 0.2$ (see text; As$^{33}$ for $\lambda$ is not shown
since the ratio of $G_{\rm obs-\sigma}$=11.2 and $G_{\rm obs-\lambda}$=4.9 
is too large for plotting). Horizontal positions for some
 metals are slightly shifted to avoid overlapping. Numerals in the
 top and bottom are adopted $G$-values.}
\label{f1}
\end{figure}

Further we find that in the Kittel's periodic table (K-P-table; in the back cover of the book), not
necessarily in other authors' tables, sum of numbers in the outer
electronic configuration matches the observed $G_{\rm obs}$ quite well. In
fact we find that besides $G \approx 1$ (from $G_{\rm obs} \approx 1$) for
s$^{1}$-electron atoms, $G=3$ (from $G_{\rm obs} \approx 3$) for
3-outer-electron atoms as in In$^{49}$(5s$^{2}$5p$^{1})$, and $G=5$ for
5-outer-electron atoms as in Nb$^{41}$(4d$^{4}$5s$^{1})$. Other examples
besides $G=1$, 3 and 5 are Mg$^{12}$(3s$^{2} \rightarrow G=2$),
Cr$^{24}$(3d$^{5}$4s$^{1}\rightarrow G=6$), Zn$^{30}$(4s$^{2}$
$\rightarrow G=2$), Tc$^{43}$(4d$^{5}$5s$^{2}\rightarrow G=7$;
not 4d$^{6}$5s$^{1}$ as in the K-P-table), and
Pb$^{82}$(6s$^{2}$6p$^{2}\rightarrow G=4$). These estimated $G$-values
are shown in Fig. 1 in the top and bottom. While the K-P-table
actually lists as 3d$^{10}$4s$^{2}$ for Zn, we ignore 3d$^{10}$, as we
ignore the same $n$d$^{10}$($n=3-5$) in Cu, Ag, Au, and Cd$^{48}$(5s$^{2} \rightarrow G$=2): if we added 10 to $G, \sigma$ 
would become much less than the observation, though $n$d$^{10}$ may be important for the electronic structure.  In the case of Cu, we know that 
the state density from d$^{10}$-bands
is confined below the Fermi energy and hence no contribution (see Fig. 7.12 of Ibach and L$\rm\ddot{u}$th\cite{ibach}).
We add further Al$^{13}$($G$=1, only 2p$^{1}$ is used instead of usual $G=3$
from 2s${^2}$2p$^{1}$, since $G_{\rm obs}$=1.2), Sr$^{38}$, Ta$^{73}$, W$^{74}$($G$=2, 6s$^{2}$),
Re$^{75}$, Ir$^{77}$($G$=2 from 5d$^{7}$6s$^{2}$, but not 5d$^{9}$ as in the K-P-table), and Tl$^{81}$. 
Altogether 22 metals in the K-P-table show that if one uses these G-values,
Eq.(\ref{Eq.2}) holds quite well with the scatter rms of
$\left|G_{\rm obs}-G \right|/G=\left|\sigma-\sigma_{\rm obs}\right|/\sigma_{\rm obs}=23{\%}$.

For the remaining 26 metals, we need to inspect in detail,
primarily because the periodic table itself is rather complicated. 
There seem two ways of guessing $G$. In the first method, given the
observed $G_{\rm obs}$, we force to choose configurations counted from the
highest term until the sum of electron numbers becomes closest to $G_{\rm obs}$,
that is we round off $G_{\rm obs}$ to integer such that
$\left|G-G_{\rm obs}\right|\leq 0.5$, namely $G \equiv
(G_{\rm obs})_{\rm round}$. Naturally $G/G_{\rm obs}$ becomes almost unity as seen in
Fig. 1(b)-lower. Though in principle there seems no reason to reject this first method which entirely ignores 
the electron configuration such as 3d$^{5}4s^{2}$,
we `feel uneasy' because many metals show much smaller deviations from unity than the relative differences of `non-identical experimental
$\sigma$-values' between the K-P-table and PSJ-table ($\pm 7{\%}$ for 38
metals).

We adopt then an alternative second method in this paper as shown below. We add deeper
`electron configurations' (hereafter E-config) for some elements than the K-P-table; examples are
Li$^{3}$(from 2s$^{1}$ to $\underline{1s^{2}}$2s$^{1}$, leading to
$G=3$), V$^{23}$(3d$^{3}$4s$^{2} \rightarrow
$\underline{3p}$^{6}$3d$^{3}$4s$^{2}\rightarrow G=11$), and
Bi$^{83}$(6s$^{2}$6p$^{3}\rightarrow
$\underline{5s$^{2}$5p$^{6}$5d$^{10}$}6s$^{2}$6p$^{3}\rightarrow 23$).
Here the added part is underlined. The last one Bi
a typical semimetal, may be noteworthy, where 1/$\sigma_{\rm obs}\equiv\rho_{\rm obs}$=1160n$\Omega$m$\approx1.5\rho_{\rm obs}$(La$^{57}$) at 300K. 
This is because by including enough deep
levels, it can be treated in the same way as others, and Bi
gives $\sigma \approx \sigma_{\rm obs}$ using $G_{\rm obs}=23.8$
(full E-config of Bi$^{83}$ is $[\{$Pd$^{46}\}$4f$^{14}]$5s$^{2}$5p$^{6}$5d$^{10}$6s$^{2}$6p$^{3})$. 
We could have assigned $G=21$ by excluding the first 5s$^2$, indicating non unique $G$-values for a large $G_{\rm obs}$.
A brief comment is given for semimetals near the section end.  
Adding Y$^{39}$(G=11), Zr$^{40}$(G=12), La$^{57}$(G=11), 
and Hf$^{72}$(G=10 from 5p$^{6}$5d$^{2}$6s$^{2}$, not 4f$^{14}$5d$^{2}$6s$^{2}$ as in K-P-table),
seven metals fall in this category (altogether 29 metals up to this point). 

Though it is possible to treat Sc, Ti and iron group similarly, we
introduce two rules below to obtain `much better' agreements with the
observations. We first introduce what we call (10$-x$)-rule.
We examine
Fe$^{26}$(listed as 3d$^{6}$4s$^{2}$ in the K-P-table), 
Co$^{27}$(3d$^{7}$4s$^{2})$ and Ni$^{28}$(3d$^{8}$4s$^{2})$, giving
`tentative' values of $G_{\rm tent}=8$, 9 and 10, respectively. However if we
introduce one rule that if $G_{\rm tent}>5$ is encountered in d-band where
the saturation is 10, we use $G=10-G_{\rm tent}$ as a subset of
d-band. Then $G/G_{\rm obs}$ becomes closer to unity. Namely,
$G_{\rm obs}$=(6.1, 3.8, 4.7) and new $G=$(2+4, 2+3, 2+2) are obtained for Fe, Co, and Ni, 
respectively, where the first numeral 2's come from s$^{2}$.
The result is $G/G_{\rm obs}=\sigma_{\rm obs}/\sigma$=(6/6.1, 5/3.8, 4/4.7)=(0.98, 1.3, 0.85).
On the other hand for
$G_{\rm tent}$ we would have obtained $G_{\rm tent}/G_{\rm obs}$=(1.6, 2.9, 2.6),
which we regard unsatisfactory. We applied this rule altogether to six
elements in $8^{\rm th}-10^{\rm th}$ group, including Ru$^{44}$, Rh$^{45}$,
and Os$^{76}$. 

This kind of rule is seen in atomic
spectroscopy \cite{condon,cox}, where d$^{x}$ and d$^{10-x}$ give the
same LS coupling terms such as $^{1}\rm{S},^{1}\!\rm{D},^{1}\!\rm{G},
^{3}\!P$, and $^{3}\rm{F}$ for $x=2$ together with
similar rules like p$^{6-x}$ and f$^{14-x}$, being called `equivalent' electrons for
the same set of ($n, l$). Also when the cohesive
energy of many metals was estimated, a similar kind of rule has been
utilized\cite{friedel}, which is broadly consistent with an extensive
calculation\cite{moruzzi}. The only strong reason however that we use
(10$-x$)-rule is because it gives better agreements with
experiments.

Finally we introduce what we call (1/2)-rule. When the $G_{\rm obs}$-value in a 
metal corresponds just inside of semi-closed shells of $g$=2, 6 or 10 (statistical weight), we
take 1/2 of these values to the last of the sum of E-config. This is a sort of extended $(10-x)$-rule. 
Underlying presumption is that
though these 2, 6 and 10 electrons are closely packed, there might be
weak breaks just in the middle of these, namely 1, 3 and 5. As a first
example, Pd$^{46}$(4d$^{10}$) shows $G_{\rm obs}$=5.2 for
$\sigma$, hence instead of adopting $G=10$, we adopt $G=5$ from
10/2. We applied the `1/2-rule' also for Ca$^{20}$(4s$^{2}, G$=1, $G_{\rm obs}$=0.6;
instead of usual $G=2$), and Sn$^{50}$. 
For six metals we combine the addition of deeper terms to the K-P-table and the `1/2-rule';
Sc$^{21}$(\underline{2p$^{6}$3s$^{2}${3p}$^{6}$}3d$^{1}$4s$^{2}\rightarrow
G=14$), Ti$^{22}$(\underline{2p$^{6}$3s$^{2}$3p$^{6}$}3d$^{2}$4s$^{2}\rightarrow
G=15$) and Ba$^{56}$(\underline{5p$^{6}$}6s$^{2}\rightarrow G=5$). Here
added terms are underlined, and 6 in 2p$^{6}$-bands (Sc and Ti) or
5p$^{6}$-bands (Ba) is replaced by 6/2=3. Also Be$^{4}(G$=3), Ga$^{31}(G$=8) and Sb$^{51}(G$=10) fall in this group. 
Again it is noteworthy that though Sb is a semimetal, it can be treated in the same way as the usual metal; 
namely $G$=10 for Sb$^{51}$, almost identical to $G_{\rm obs}$=10.0, comes from an addition of 5 
from K-P-table (5s$^{2}$5p$^{3}$) and 5=10/2 (1/2-rule for 4d$^{10}$)
from full E-config of Kr$^{36}$4d$^{10}$5s$^{2}$5p$^{3}$.
Note that the resistivity of Sb ($1/\sigma_{\rm obs}$=413n$\Omega$m) is similar to that of Sc, Ti, or Zr.
Further, another semimetal As$^{33}$ (Ar$^{18}$3d$^{10}$4s$^{2}$4p$^{3}$) shows 
$G_{\rm obs}$=11.2 (1/$\sigma_{\rm obs}$=333n$\Omega$m from PSJ-Table\cite{psj}) so that
$G$=10 (1/2-rule for 3d$^{10}$) may be appropriate. 
 
Although use of `(1/2)-rule', so far applied to ten metals, is due primarily to better fit the observations, we want to
stress that without this rule $\left|G-G_{\rm obs}\right| /G_{\rm obs}$ would
become much larger than in other metals in the same group where these
rules are not needed (see Fig. 2).

The remaining four metals are problematic in one way or another: 
Mn$^{25}(G_{\rm obs}$=83$>$25 of atomic number, but see Fig.26 of Meaden \cite{meaden}),
Mo$^{42}(G_{\rm obs}$=2.5; $G$=1, though 5 is possible), Pt$^{78}(G_{\rm obs}$=5.0, $G$= 9, though
$G$=2 is possible), and Hg$^{80}$(liquid; $G_{\rm obs}$=28.5 and
$G$=27). 

Figure 1 (b)-upper shows, excluding the problematic four metals above, 
an rms scatter of $\pm 20$\% which is larger than the observation error of
(rms)$_{\rm obs}=\pm 7$\% mentioned before. This suggests that the scatter
stems largely from not-yet incorporated causes [(rms)$_{\rm obs}=
\pm 14$\% for $G=(G_{\rm obs})_{\rm round}$ in Fig. 1(b)-lower]. 

For the thermal conductivity $\lambda$, we also show $G_{\rm obs-\lambda} \equiv \lambda_{1}/\lambda_{\rm obs}$ [Fig. 1(a)] and
$\lambda/\lambda_{\rm obs}$ [Fig. 1(b)-upper] in thick marks
$[\lambda_{1} \equiv \lambda(G=1$)].  
Here we employed Eq.(\ref{Eq.2}) for $\lambda$ with the same $G$ used for $\sigma$. We plotted only nine
metals showing large departure from the Wiedemann-Franz law\cite{kittel,
ibach,ashcroft} ($\left|\,C_{\rm WF}-1 \right| \geq 0.2$): Ti, Cr, Co, As, Y, Zr, Sb, W, and Bi.
Here $C_{\rm WF}\equiv\lambda/(\sigma TL_{\rm z})$, where $L_{\rm z}\equiv \pi^{2}k_{\rm B}^{2}/3e^{2}$. 
We find no appreciable differences from
$G_{\rm obs}=\sigma_{1}/\sigma_{\rm obs}$ even for those metals of large
$\left|\,C_{\rm WF}\!-\!1\right|$ (though in As it is rather large).

Figure 2, which is supplementary to Fig. 1, shows that the
position of estimated $G$'s in E-config. For example, E-config of
Fe$^{26}$ is Ar$^{18}$3d$^{6}$4s$^{2}$ and we adopted $G=2+4=6$ (4
comes from $10-6$ in d$^{6})$ added from outer ones, where 4 is within
the 3d-band. Hence for Fe we plotted at 3d. Fig. 2 shows rather
systematic behavior, particularly among 4-6 periods (starting
from K, Rb, and Cs). This indicates that our choice of $G$, though
adopted only to match the observations, appears to be rooted from some
physical basis. In fact we notice that groups 3-4 (Sc$^{21}$,
Ti$^{22}$, Y$^{39}$, Zr$^{40}$, and La$^{57})$, which stem deeper
configurations than other metals, all have configurations where
electrons are filled in the outer bands before inner bands become
filled up or closed;
e.g. Sc$^{21}$=\{Be$^{4}$\}2p$^{6}$3s$^{2}$3p$^{6}$\underline{3d}$^{1}$\underline{4s}$^{2}$
instead of \{Be$^{4}$\}2p$^{6}$3s$^{2}$3p$^{6}$\underline{3d}$^{3}$,
while the 3d-orbit only saturates at Cu$^{29}$ as 3d$^{10}$. This is of
course typical characteristics of the earlier transition elements. It
might suggest some unstableness of so to speak \textit{heavier upper
floors} than e.g. noble metals, and as a consequence involvement of
deeper orbits. 

\begin{figure}
\begin{center}
\includegraphics[width=7.5cm]{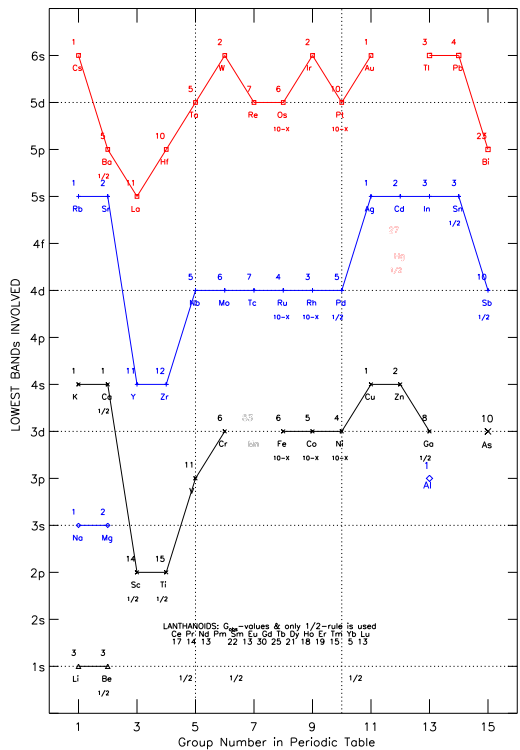}
\end{center}
\caption{(Color online) The lowest positions in the electron configuration where the
 final count is made to fix $G$ from upper levels vs. group
 number. Numbers above metal names are $G$-values adopted (same as in
 Fig. 1) and below names are whether (10$-x$) or (1/2) rule is
 employed.}
\label{f2}
\end{figure}

In addition to 49 metals already discussed, we show $G_{\rm obs}$-values for rare earth metals in Fig. 2 
and find $\sigma / \sigma_{\rm obs}
\approx 1$, mainly because $G_{\rm obs}\ge10$, except for Yb$^{70}$, so that it is easier to find $G/G_{\rm obs}\approx 1$ 
(rms=$\pm 17\%$ for 13 metals and the lowest orbits are in 4d, 4f, and 5p; if
Tm$^{69}(G/G_{\rm obs}=1.53)$ is excluded rms becomes $\pm 12\%$).
A conclusion from Fig. 2 is that our choice of $G$-values shows rather systematic
distribution among E-config for various metals, supporting the choice,
if not prove, besides giving nearly correct values of $\sigma_{\rm obs}$.

This and the next paragraphs inspect semimetals. It is believed that the number of conducting electrons (and holes)
per atom is much smaller than in ordinary metals by factors of 10$^{-2} $ (As), 10$^{-3}$ (Sb), and 10$^{-5}$ (Bi) 
(e.g. Chap.8, table 7 in Kittel\cite{kittel}, Chap. 14.2 in Abrikosov\cite{abrikosov}).
However, for (As, Sb, Bi), $1/\sigma_{\rm obs}$ =(333, 413, 1160)n$\Omega$m are close to those for  (Hf, Zr,  1.5$\times$La) at $\sim$300K, respectively. 
Also the departure from the Wiedeman-Franz law, though rather large, is generally similar to those for Ti, Cr, Co, and others as seen in
Fig.1b-upper. Further, observations (e.g. Fig.5 of Issi\cite{issi}) show roughly $\sigma\propto T^{-1}$ around $T_{\rm room}.$ 
Besides, for the mass densiy, melting point, bulk modulus, and cohesive energy,
we can find metals showing similar values. The only difference seems to be
the lattice structure of `rhomb' of semimetals in Kittel's tabulation.

The very small density $n_{\rm e}$ (e.g. observed from the Hall effect in the magnetic field of $B=10^{4}$G) 
is expressed\cite{issi,gonze} as $n_{\rm e}(B\neq 0)/n=(\Delta E_{\rm e}/E_{\rm F})^{\rm{1/2}}  {\bf det}\:\:{\bf m} \ll 1$ in unit of the total electron density $n$.
Here ${\bf m}$ is the diagonal mass tensor in $m$ unit, and  
$\Delta E_{\rm e}$ is the difference between the Fermi energy 
and extremum energy for electrons. Both factors, ``the determinant of  $\bf{m}"$ and $(\Delta E_{\rm e}/E_{\rm F})^{1/2}$, 
are known much smaller than unity.
Therefore  as far as $\sigma$ and $\lambda$ at $T_{\rm room}$ without magnetic fields are concerned, they behave 
as if ${\bf det}\,{\bf m}=1\,(m^{\ast}=m)$ 
and $\Delta E_{\rm e}/E_{\rm F}=1$ were to hold, and hence
in this empirical paper, it may be allowed to say that semimetals
can be treated similarly as ordinary metals.  

Below is an extra note for the semiconductor. It is well-known that 
the thermal conductivities of Si$^{14}$ and Ge$^{32}$ are not much different from ordinary metals,
unlike much reduced electrical conductivity due to the gap of $\sim$1eV; namely 
$\lambda_{\rm obs}$(Wm$^{-1}$K$^{-1}$)=148(Si)$\approx$ 147(Ir) or $\lambda_{\rm obs}$=
60(Ge)$\approx$ 58(Rb) at 300K. Thus assigning similarly $G$ to them, we find that Si (Ne$^{10}$3s$^2$3p$^2$) shows  $G_{\rm obs-\lambda}$=1.78 and
$G=2$ from 2p$^2$, giving $\lambda/\lambda_{\rm obs}$=0.89, and Ge(Ar$^{18}$3d$^{10}$4s$^2$4p$^2$) shows $G_{\rm obs-\lambda}$=3.90 and
$G=4$ from 4s$^2$4p$^2$, giving $\lambda/\lambda_{\rm obs}$=0.98; satisfactory results in selecting $G$ by reasonable E-config too.\\ 

\begin{center}
\bf{4.} Z=1 is Expected in All Metals for $\sigma$
\end{center}

Though Eq.(2) needs only values of $n_{\rm atom}$, naturally we wish to know
$n$ or $Z\equiv n/n_{\rm atom}$. This $Z$ value is tabulated only for 
20 metals in Kittel's table \cite{kittel}, while the other quantities are
tabulated for almost all metals, indicating that to assign $Z$-value is not easy. 
This means in turn that there is no reliable theory how to assign $Z$ for electrical conductivity, so that
we search empirically for $Z$ in multivalent metals. Below we use the letter
$`Z_{\sigma}'$, emphasizing so-called valence electron numbers only applicable
to electrical conductivity of elemental metals $\sigma$, i.e. not for alloys. We also assume that $Z_{\sigma}$ is equal
for thermal conductivity and possibly at all temperature range.

The result of Sect.2 indicates that there seems
no room for $Z_{\sigma}\neq 1$, because $\sigma=Z_{\sigma}\times n_{\rm atom}e^{2}\tau_{0}/(mG)$ with $\tau_{0}=\hbar/(k_{\rm B}T)$ and
$G$ listed in Figs. 1-2 
cannot reproduce the observations. In other words, the fact that we can
reproduce the observation quite well without knowing $Z_{\sigma}-$value suggests by itself already
$Z_{\sigma}=1$. Namely 
\begin{equation}
Z_{\sigma}\equiv n/n_{\rm atom}=1       \label{zeq1}
\end{equation}
is expected for the majority of elemental metals in electrical and thermal conductivities,
if not for e.g. cohesive energy and compressibility, or if not for alloys.
[Note that $Z_{\sigma}$ and $G$ are different physical quantities. The $G$ is the number of bands `responsible' for $\sigma$ 
for `a single electron', which is only one per atom in the elemental metal
as expressed $Z_{\sigma}=1$ (here the word `responsible' means
that the state density $D(E)$ for that band shows non-zero values beyond the Fermi energy; see Appendix).]

Nevertheless we inspect $Z$-values using two options below.
In both cases we assume $\tau_{i}=\tau_{0}=\hbar/k_{\rm B}T$ of Eq.(1)
for each $i$-band, and the sum of the number of
band is equal to $G$, namely $\Sigma_{i=1}^{G}1=G$.
Let us first assume that the
total resistivity for multi-band metals is given by
\begin{equation}
\rho\equiv 1/\sigma=\Sigma\rho_{i}=\Sigma\frac{m}{e^2n\tau_{i}}
=\frac{m}{e^2n}\frac{G}{\tau_{0}}
 \label {rho_sum}
\end{equation}
 as in Matthiessen's rule (e.g. Eq.(16.22) of Ashcroft and Mermin\cite{ashcroft};
of course `without impurity').
Here $\rho_{i}$ is defined as $\rho_{i}=m/(e^2n\tau_{i})$, and   
use is made of $\Sigma (1/\tau_{i})=(\Sigma_{i=1}^{G}{1})/\tau_{0}=G/\tau_{0}$.
In order to obtain an equality between Eq.(5) and Eq.(2) which uses $n_{\rm atom}, n=n_{\rm atom}$ is required as in Eq.(\ref{zeq1}).
Thus, although the assumption that every band takes the same $\tau_{0}$ might seem
a very crude assumption, we should remember that $G$-values were guessed by assuming that every band
has equally `one' contribution each to $G$ with the same $\tau_{0}$, resulting in good agreement with observations. 

As an alternative option, if we assume $\sigma=\Sigma\sigma_{\rm i}=\frac{e^2n\tau_{0}}{m}\times G$,
as in Eq.(13.24) of Ashcroft and Mermin\cite{ashcroft}
again with $\tau_{i}=\tau_{0}$ for all $i$, we obtain
\begin{equation}
\sigma=\Sigma_{\rm i}\sigma_{\rm i}=\frac{e^2n_{\rm atom}\tau_{0}}{m}\times GZ_{\sigma}=\sigma_{\rm Eq.(2)}\times G^2Z_{\sigma}.  \label {sigma_sigma}
\end{equation}
This is compatible with Eq.(2) only when $Z_{\sigma}=1$ and $G=1$
since $G\ge 1$ and $Z_{\sigma}\ge 1$. Hence
we discard this latter option. 

Note that only Aschcroft and Mermin\cite{ashcroft} in p.250 mention, without
any theoretical explanation,
on the additive current density and hence $\sigma=\Sigma_{i}\sigma_{i}$, but all the other text books
cited in this paper and some other books did not say anything on this important issue.
On the other hand, the Mattiessen rule is mentioned in many books, expressing
that it is applicable if two (or more) distinguishable sorces of scattering 
as in $\rho=\Sigma_{i}\rho_{i}$, but it is nowhere mentioned that `distinguishable sorces' should not include different
electron bands. Thus in this empirical paper we adopted Eq.(\ref{rho_sum}), which
is consistent to observations, though $Z$-values are not needed 
in Eqs.(1-2) (see theoretical derivation of Eq.(\ref{bohm})
supporting $Z\approx 1$).

We look these results of Eqs.(\ref{zeq1}) and (\ref{rho_sum}) from the macroscopic stand point of view. The classical equation of motion for a one electron is
$mdV_{\rm x}/dt=-eE_{\rm x}-mV_{\rm drift}\nu$, where $E_{\rm x}$ is the imposed electric field 
and $V_{\rm drift}$ is the effective drift velocity both in $x$-direction and $\nu$ is the number of collisions in unit time.
Then  the option we adopt means that the total resistive force for one electron
 $-mV_{\rm drift}\,\nu$ is the summed resistive force
of each band in elemental metals, which we consider as mutually independent.
Namely as Eq.(2) or (5), the sum of each band means $\nu=\sum \nu_{i}
=\sum(1/\tau_{i})=G/\tau_{0}$ and $Z_{\sigma}$=1, e.g.
 either for $G=3$ in In$^{49}$(5s${^2}$5p$^{1}$) or $G=4$ in Pb$^{82}${(6s$^{2}$6d$^{2}$) with the same $\tau_{0}$.
Also even if one doubts (10-$x$)-rule or 1/2-rule, the result for $G$ without these rules is not worse than a factor of about 2 for
those metals which employed either one of the two rules.

The conclusion of this section is simply $Z_{\sigma}=1$ or $n=n_{\rm atom}$,
namely Eq.(\ref{zeq1}). This means that the number of electrons per one atom is effectively only one,
which is responsible for $\sigma$ when given the electric fields and $\lambda$ when given the temperature gradient,
but $Z_{\sigma}=1$ does not necessarily apply for other physical quantities such as cohesive energy or with imposed magnetic fields.
Further, $Z_{\sigma}=1$ is mutually consistent with the condition that every $\nu_{i}\equiv1/\tau_{i}$ is the same as $1/\tau_{0}$.
The total collision frequency is $\nu_{\rm tot}=\Sigma_{i=1}^{G}\nu_{i}=G/\tau_{0}$. Here $\nu_{\rm tot}$ acts in
the resistive force for `a single
electron' in the classical equation of motion as $-mV_{\rm drift}\nu_{\rm tot}=-mV_{\rm drift}G/\tau_{0}.$\\

\begin{center}
\bf{5. Effective Mass $\simeq$ True Electron Mass for $\sigma$}
\end{center}

This section inspects if the use of the true electron mass $m$ 
in Eq.(2) is allowed rather than the effective electron mass $m^{*}$,
even though Eq.(2) satisfies observations quite well.  
For this purpose we use electron specific heat at constant volume 
$C_{\rm Ve}$, which is only `discernible' below a few Kelvin (of course
$C_{\rm Ve}$ is not zero even at high temperature).
The ratio of the observed $C_{\rm Ve-obs}$ 
to 
\begin{equation}
C_{\rm Ve-free}=\frac{\pi^{2}}{3}D(E_{\rm F})_{\rm free}k_{\rm B}^{2}T
=\frac{\pi^2}{2}\frac{k_{\rm B}^{2}T}{E_{\rm F}}
\label{Cv1}
\end{equation}
from the free electron model
is conventionally expressed as $m^{\ast}/m$ ($m^{\ast}$ is
called the thermal effective electron mass \cite{kittel}).
Here $D(E_{\rm F})_{\rm free}=3/2E_{\rm F}$ is the density of states/atom for the free electron model.
This comes from $C_{\rm Ve-free}\propto E_{\rm F}^{-1}\propto m$ and similarly we designate $m^{\ast}$ from $C_{\rm Ve-obs}\propto m^{\ast}$.
The ratio $m^{\ast}/m=C_{\rm Ve-obs}/C_{\rm Ve-free}$ amounts to ten or more for the transition metals
\cite {ashcroft,ibach, kittel}. 


Extensive numerical calculations by Moruzzi et al.\cite{moruzzi}
for non-free electron models (the local density theory and the 
`muffin-tin' model) tabulate, besides the calculated Fermi energy, 
the density of states $D(E_{\rm F})$ at the Fermi energy for each metal.
Using the latter expressed as $D(E_{\rm F})_{\rm Morz}$ , we find  
\begin{equation}
\frac{m^{\ast}}{m}=\frac{C_{\rm Ve-obs}}{C_{\rm Ve-theory}}
=\frac{C_{\rm Ve-obs}}{\pi^2D(E_{\rm F})_{\rm Morz}k_{\rm_B}^2T/3}=1.05\pm 0.29 \label{mm-ratio}
\end{equation}
for 24 metals ($C_{\rm Ve-obs}$ from Kittel\cite{kittel}).
Below shows some indication of the accuracy of $m^{\ast}/m$ in the theory\cite{moruzzi},
other than the observational errors. 
Six metals not included in the above 24 metals show $m^{\ast}/m>2.0,$
where the values of $m^{\ast}/m$ are 4.9(Sr), 3.1(Y), 2.6(Sc), 2.5(Mn), 2.4(V), and Nb(2.4).
For example, $D(E)$ plotted against $E$ for Sr shows that $E_{\rm F}$ is in the bottom of a sharp valley \cite{moruzzi} 
suggesting that $D(E_{\rm F})=0.31$states/eV$\cdot$atom could have been larger,
and accordingly $m^{\ast}/m$ may become closer to unity, if the theory
becomes further refined (cf. $D($\EF)=0.45 in simpler sodium). A similar situation
is seen in Y [$D(E_{\rm F})$=1.41], which has a steep peak of $D(E)$ just below $E_{\rm F}$. 

Taking into account of errors from theory and observation, this result of $m^{\ast}\approx m$ supports to 
use the true electron mass $m$ in Eq.(2) (no magnetic fields), besides Eqs.(1) and (2)
reproduce the observations quite well. Note that the departure from unity of $\sigma/\sigma_{\rm obs}$ in Fig.1 (b)-upper
does not seem to be correlated with $m^{\ast}/m$, nor with the lattice structures such as fcc or bcc.

The following will clarify that $m^{\ast}/m$ from $C_{\rm Ve}$ should be the same
as $m^{\ast}/m$ to be used in $\sigma$ at room temperature range. 
If one adopts the classical thermal conductivity 
$\lambda_{\rm classical}=\frac{1}{3}C_{\rm V}L_{\rm mfp}V\times n$, 
we recover Eq.(2) for $\lambda$ multiplied by $G$, inserting $C_{\rm V}=C_{\rm Ve-free}=
(\pi k_{\rm B})^{2}T/(2E_{\rm F})=(\pi k_{\rm B})^{2}T/(mV_{\rm F}^{2}),
L_{\rm mp}V=V_{\rm F}^{2}\tau_{0}$, and $n=n_{\rm atom}$ from Eq.(\ref{zeq1}) ($L_{\rm mfp}$ is the mean free path). The classical
form $\lambda_{\rm classical}$ is now considered to be applicable to a single band, $G=1$, or rather
if we change the definition from $\lambda_{\rm classical}$ to $\lambda_{\rm semi-classical}=\frac{1}{3}C_{\rm V}L_{\rm mfp}Vn/G$, then
we recover exactly Eq.(2) for $\lambda$.
Since $\lambda_{\rm semi-classical}\propto C_{\rm Ve}\propto E_{\rm F}^{-1}\propto m^{\ast}$, 
this shows that use of $m^{\ast}/m\approx 1$ is supported also in
$\lambda$, which in turn should use the same electron mass in $\sigma$ of Eq.(2).

The conclusion of this section is the same as the section title,
though one might add `which is found using $C_{\rm Ve-obs}
/C_{\rm Ve-theory}$'\\


\begin{center}
\bf{6. Reducing Bardeen's $\tau$ to $\tau_{0}$}
\end{center}

This section will relate our $\tau_{0}=\hbar/k_{\rm B}T$ to the collision time of Bardeen\cite{bardeen}
$\tau_{\rm B}$ used for $\sigma=e^{2}n\tau_{\rm B}/m$ in his Eq.(6)$\equiv$Eq.(bd6), 
which is applicable to the monovalent metals at $T_{\rm room}$ range. Eq.(bd52) is
\begin{equation}
1/\tau_{\rm B}=\hbar\pi^{3}nk_{\rm m}^{-2}(dk/dE)_{\rm m}(T/Mk_{\rm B}\Theta^{2})C^{2}. \label{bardeen0}
\end{equation}
This\,($\hbar^{2}$ should read $\hbar$ as above) is converted to 
\begin {equation}
\frac{\tau_{\rm B}}{\tau_{0}}
  =\frac{3M}{\pi m}\left(\frac{k_{\rm B}\Theta}{C}\right)^{2}
  =\frac{4}{2^{1/3}\pi}\frac{3MC_{\rm s}^{2}}{E_{\rm F}}\frac{E_{\rm F}^{2}}{{C}^{2}}
  =1.01\left(\frac{C}{E_{\rm F}}\right)^{-2}. \label{bardeen} 
\end {equation}
The first equality used $k_{\rm m}(dE/dk)_{\rm m}$=2\EF =
$\hbar^{2}k_{\rm m}^{2}/m$ with $n=k_{\rm m}^{3}/3\pi^{2}$.
The subscript $m$ refers to maximum values. This first equality is
equivalent to Eq.(12.3) of $1/\tau$ in Jones's review\cite{jones}. 
The second equality uses \kB$\Theta=C_{\rm s}\hbar q_{\rm m}=2^{1/3}C_{\rm s}\hbar k_{\rm m}$ [Eq.(bd11)], where $C_{\rm s}$ is the sound speed,
$q_{\rm m}=2^{1/3}k_{\rm m}$ is from $n=$\natom$=q_{\rm m}^{3}/6\pi^{2}$,
 and $C$ is the distorted potential (interaction constant of Sommerfeld and Bethe\cite{sommerfeld}).
The final equality
adopts the relation (see below)
\begin{equation}
3MC_{\rm s}^{2}\approx E_{\rm F}, \label{bohm}
\end{equation}
for atom of mass $M$.
Then if $(C/E_{\rm F})^{2}=1$ for all monovalent metals, we obtain $\tau_{\rm B}\approx\hbar/k_{\rm B}T\equiv \tau_{0}.$
Bardeen\cite {bardeen}(p.697) may ``have underestimated  the Umklapp processes so that true values of $C/E_{\rm F}$ 
should be larger by perhaps ten or fifteen percent" than the average value of 0.8
of his theory for 7 monovalent metals. Namely he implies $\tau_{\rm B}/\tau_{0}=1.2\sim 1.3$,
while for the same metals (Fig.1b-upper) $\tau_{\rm obs}/\tau_{0}=\sigma_{\rm obs}/\sigma$
=1.1$\pm 0.2$. [See also the original defining Eq.(34.24) of $C$ as ``the mean kinetic energy of
 conducting electrons" in Sommerfeld and Bethe\cite{sommerfeld}, their Eq.(34.31) above, and Sect. 36 c) p.191.]

We empirically find below that Eq.(\ref{bohm}) holds if we assume $C_{\rm s}^{2}=C_{\rm tot}^{2}$, where
$C_{\rm tot}^{-3}\equiv C_{\rm long}^{-3}+2C_{\rm trans}^{-3}$ as in Eq.(5.18) of Ibach and L$\rm \ddot{u}$th\cite{ibach}
for the total of longitudinal and transverse waves. 
Namely $3MC_{\rm tot}^{2}/E_{\rm F}^{\rm Morz}=0.93\pm0.20$ is 
obtained for 13 metals of Be, Mg, Al, Cr, Fe, Ni, Cu, Zn, Pd, Ag, and Cd, and further
Ti and Zr  (all metals with tabulated $C_{\rm long}$ and $C_{\rm trans}$  values\cite{rika}) by use of 
the theoretical $E_{\rm F}^{\rm Morz}$ by Moruzzi et. al\cite{moruzzi}, 
while if one uses \EF$^{\rm free}$ we find $3MC_{\rm tot}^{2}/E_{\rm F}^{\rm free}=1.22.$ 
Similar values are obtained from Papaconstantopoulos\cite{papaconst}, where
the calculated Fermi energy is still somewhat different from
Moruzzi et al.\cite{moruzzi} even in Na; namely \EF/\EF$^{\rm free}$=0.90 in the former and 1.07
in the latter. [There seems no numerical correlation between $3MC_{\rm tot}^{2}/E_{\rm F}^{\rm Morz}$ and $\sigma/\sigma_{\rm obs}$,
which might have related to e.g. deformation potentials $\propto C/E_{\rm F}$.] 

Likewise if we use the tabulated slender `rod' sound speeds \cite{rika}, 
we find  $MC_{\rm rod}^{2}/E_{\rm F}=1.26\pm0.24$ using theoretical \EF
\,\,for 11 metals without Ti and Zr (no $C_{\rm rod}$ data), while
$MC_{\rm rod}^{2}/E_{\rm F}^{\rm free}=1.78$. This may be a one dimensional
version of Eq.(\ref{bohm}). Since Eq.(\ref{bohm}) is independent whether external electric fields exist or not, 
$C_{\rm tot}$ seems more appropriate than $C_{\rm rod}$.

Theoretically Eq.(\ref{bohm}) can `roughly' be derived as follows. We convert 
$C_{\rm tot}^{2}=\omega^{2}/k^{2}$ by using $\omega^{2}=\Omega_{\rm i}^{2}/\epsilon=\Omega_{\rm i}^{2}k^{2}/(k^{2}+k_{0}^{2})$ and
ion plasma frequency $\Omega_{\rm i}^{2}=4\pi e^{2}nZ/M$. Here
$\epsilon(k)$ is the dielectric constant 
and $k_{0}$ is the Thomas-Fermi wave number defined by 
$k_{0}^{2}=4\pi e^{2}\partial n/\partial \mu$ ($\mu$=chemical potential$\approx$\EF). If one uses the state density at the Fermi energy
for the free electron model, $\partial n/\partial \mu=3n/2E_{\rm F}$, 
the Bohm-Staver relation $3MC_{\rm tot}^{2}=2ZE_{\rm F}k_{0}^{2}/(k^{2}+k_{0}^{2})=2ZE_{\rm F}$ is obtained
for $k\rightarrow 0$ [\,e.g. Eqs.(26.2-8) of Ashcroft and Mermin\cite{ashcroft}].
This holds even if we replace $M\Omega_{\rm i}^{2}$ by 
$M\Omega_{\rm i}^{2}-k^{2}V'$ from the `shallow' constant potential
$V'(k)$ in the ion core (see e.g. Eq.(6.92) and Fig.88(c) of the screened pseudo-potential in Ziman\cite{ziman72}).
Incidentally Ziman notes that the formula for
the screened potential, i.e. his Eq.(6.93)[=(6.92)/$\epsilon(k)$] ``is the formula derived by Bardeen
in 1937 for the matrix elements of the electron-phonon interaction...".  On the other hand
if $k\rightarrow k_{0}$ and $Z$=1 are assumed, we recover exactly Eq.(\ref{bohm}), which is also valid
if the core potential part is negligible such as in the often-used-Ashcroft model of $V'=0$\cite{rossiter}.
(The Thomas-Fermi screening is the spherically symmetric Fourier transform of the screened electron
potential $\phi(r)=-Ze^{2}{\rm exp}(-k_{0}r)/4\pi\epsilon_{0}r$.)
Equation(\ref{bohm}) implies
that ion vibrations can be expressed by the electron properties or motions in (nearly) thermal
equilibrium. 

The conclusion of this section is that Bardeen's $\tau_{\rm B}$
is found to be equal to $\tau_{0}=\hbar/k_{\rm B}T$ of Eq.(1) within
10 to 20$\%$ accuracy using primarily the empirical relation Eq.(\ref{bohm}).
Conversely if one adopts the Bardeen theory, Eq.(1) is theoretically derived with $G=1$ in Eq.(2) within the same accuracy.
 As for the absolute values of the electrical resistivity in the non-monovalent metals,
$G$-values, i.e. the number of responsible
electron bands take care also at $T_{\rm room}$ range. 
\\

\begin{center}
\bf{7. Collison Time of Eq.(1) from Much Reduced Effective Electron Density}
\end{center}
The very fact that the Bardeen 
$\tau_{\rm B}$ can, in the end, be expressed without any parameters related to ions as shown in Sect.6 
leads to a conjecture that the theory might be constructed without
 `formally' introducing phonons, which is treated in this and the following sections.
For $T\ll T_{\rm room}$ we briefly discuss in Sect.9. 


As is well-known \mfp=\VF$\tau$  becomes 40nm in Cu for example, if one uses $\tau$ from the observed electrical resistivity, or
from Eq.(1) giving $2.55\times10^{-14}$s at 300K.
This $L_{\rm mfp}$ is very much larger than the
mean atomic distance of $d\equiv n_{\rm atom}^{-1/3}=n^{-1/3},$
e.g. $d$(Cu)=0.23nm. This does not
mean that the collision cross-section $S$
is many order of magnitudes smaller than $d^{2}=(0.23$nm)$^2$.
(If we adopt the observationally supported
Eq.(1), $S\simeq d^{2}/2$ is derived as shown below.)
But in fact only a small fraction of
$\alpha k_{\rm B}T/E_{\rm F}$ is able to collide elastically 
because the Fermi distribution function is saturated at unity below \EF
\; ($\alpha\approx$ 3 or so,
see below;\;\kB T/\EF=3.7$\times10^{-3}$ for Cu).
 
We then adopt the following simple form
\begin{equation}
1/\tau=n_{\rm eff}V_{\rm F}S, \label{cross-section1}
\end{equation}
with the defining equations of
\begin{equation}
n_{\rm eff}=\frac{\alpha k_{\rm B}T}{E_{\rm F}}n\quad\quad\quad S=\alpha_{\rm s}d^{2}. \label{cross-section2}
\end{equation}
Here $S$[m$^{2}$] is the ionic cross-section with a
parameter $\alpha_{\rm s}$ for an assumed square shape
area $d^{2}$.
\it We stress that the form of Eqs.(\ref{cross-section1}-\ref{cross-section2})  is 
expected already in Eq.(\ref{bardeen0}) from Bardeen \rm because $1/\tau_{\rm B}\propto nk_{\rm m}^{-2}T\propto nST.$
Using the constant electron speed of \VF,
Eq.(\ref {cross-section1}) immediately yields
\begin{equation}
\tau=\frac{E_{\rm F}}{\alpha k_{\rm B}T}\frac{1}{nV_{\rm F}S}
              =\frac{1}{\alpha k_{\rm B}T}\frac{\hbar k_{\rm F}}{2}\frac{(3\pi^{2})^{1/3}}{\alpha_{\rm s}k_{\rm F}}
              =\frac{3.09}{2\alpha\alpha_{\rm s}}\frac{\hbar}{k_{\rm B}T}, \label{tau-cross-section}
\end{equation}
where the second equality used
\EF/\VF=$\hbar$\kF/2 and $1/nS=(3\pi^2)^{1/3}/\alpha_{\rm s}$\kF\, from $n$=\kF$^{3}/3\pi^{2}$ and $d=n^{-1/3}$.

We estimate $\alpha$ and $\alpha_{\rm S}$ below. Regarding $\alpha$,
though it is often stated $\alpha =3\sim4$ from the rough inspection of the steeply decreasing part
of the Fermi-distribution as compared with the almost constant part, we wish to be more specific
about $\alpha$. The electron specific heat given by
$C_{\rm Ve-free}=\frac{\pi^{2}}{2}k_{\rm_B}^2T/$\EF of Eq.(\ref{Cv1}) for the free electron model,
which may be applicable to alkali and noble metals, can be rewritten as 

\begin{equation}
C_{\rm Ve-free}=\frac{3k_{\rm B}}{2}\frac{\alpha k_{\rm B}T}{E_{\rm F}}
                =C_{\rm Ve-all}\frac{n_{\rm eff}}{n}\quad\quad \alpha=\pi^{2}/3.        \label{Cv2}
\end{equation}
This relation is usually interpreted as equal to $C_{\rm Ve-all}$=3\kB/2 for all electrons
(freedom of 3 in unit energy \kB$T$/2, in contrast to freedom of 6 in lattice) multiplied by a fraction of the responsible electron number density due to 
the Fermi distribution. Therefore we adopt this $\alpha=\pi^{2}/3\simeq 3.3$. 
The origin\cite{ashcroft, kittel} of the value $\pi^{2}/3$ is as follows:
$\pi^{2}/3=\int_{-\infty}^{\infty}x^{2}(1-f)fdx$, where $f$ is the Fermi distribution function for the energy $\epsilon$,
$f=1/({\rm exp}(x)+1), x=(\epsilon-E_{\rm F})/k_{\rm B}T$ and $-df/dx=(1-f)f$ (very close to a Gaussian). Since $\int_{-\infty}^{\infty}(1-f)fdx=1$,  
we can regard this $\alpha$ as equal to the second (lowest non-zero) moment of 
$(1-f)f$, $<x^{2}>_{(1-f)f}=\pi^{2}/3=\alpha$, 
or the spread of $(1-f)f$ in unit of \kB$T$,
hence $\alpha$ has a clear physical meaning.

 We turn to the estimation of the upper and lower limit of the cross-section $S$, i.e. $\alpha_{\rm s}$.
The upper limit is $S\leq d^{2}=n^{-2/3}$, namely $\alpha_{\rm s}=1$, because
in this expression the space around any ion core is filled up, and
no electrons can freely pass. This is true for
any of the 3-D Bravais lattice structures, or for different forms of $S$
such as $S_{\rm max}=d^{2}\equiv\pi r_{\rm d}^{2}$
The lower limit is somewhat indeterminate, but we take a rather safer Bohr value of $S_{\rm min}=\pi r_{0}^{2}$.
The cross section $S_{\rm Eq.(1)}$ expected from Eq.(1) gives 
is in fact $S_{\rm Eq.(1)}=0.47d^{2}\approx d^{2}/2$\,(or $\alpha_{\rm s}\approx 1/2)$,
where use is made of Eq.(\ref{tau-cross-section}) and $\alpha=3.3$ in Eq.(\ref {Cv2}). 

Thus we reach 
\begin{equation}
\pi r_{0}^{2} \le S_{\rm Eq.(1)}\approx d^{2}/2 < d^{2}. \label{S-range}
\end{equation}
Here e.g. in Cu, $\pi r_{0}^{2}=0.170d^{2}$ and $d^{2}/2=2.94\pi a_{0}^{2}$.
Therefore without using Eq.(1), expected cross sections fall rather
narrow ranges of $0.17d^{2}<S<d^{2}$ (Cu) and it may well be so for the majority of elemental metals at $T_{\rm room}$-range
as long as $m^{*}\approx m$ and $Z_{\sigma}\approx 1$ hold as shown in Sects.4 and 5. 
The method of this section at the `present form without further elaboration' cannot be more accurate than this, though the simplicity
may count.

We call attention again to the fact that any concept of phonon or lattice vibration were not needed in the above.
In fact, in the expression of $\tau\propto$\hbarkbt in Eq.(\ref{tau-cross-section}), 
$`\hbar'$ arises from
`\EF$/V_{\rm F}' \propto \hbar$ concerning electrons,
and $`k_{\rm B}T'$ is from $n_{\rm eff}\propto T$, which originally came
from the steeply decreasing part of the `electron' Fermi distribution.

In many textbooks, it is argued that the only
cause of the non-zero resistivity stems from the ion vibration
(at the room temperature range). But since the `ion' vibration can be
expressed by Eq.(\ref{bohm}) in terms of the (`electron') Fermi energy,
we may as well express the electrical conductivity in terms of electron
behavior.

Thus conversely, once the cross-section of order of $S\approx d^2/2$ is accepted, we can conclude that
$\tau_{0}$ is on the order of $\hbar /k_{\rm B}T$, and that the precise value
is supported as $\tau_{0}=\hbar /k_{\rm B}T$ from the observations as in Sect.2 
for elementary metals at $T_{\rm room}$. Since the electron half-wavelength at the Fermi energy is $\lambda_{\rm F}/2=\pi/k_{\rm F}=(\pi/3)^{1/3}n^{-1/3}=1.016d,$ 
$S=(\lambda_{\rm F}/2)^{2}/2$ may be used as an alternative expression for $S$.

We close this section by inspecting how to interpret $n_{\rm eff}/n$
in terms of the Bardeen theory.
If \EF\, which used \EF$^{\rm {free}}$ above is different from the
true \EF, we must multiply \EF/\EF$^{\rm{free}}\times 2/(d$log$E/d$log$k)_{\rm m}$.
Ignoring this factor and $(C/E_{\rm F})^{2}$, we find that Eq.(\ref{bardeen0}) is physically equivalent to Eq.(\ref{cross-section1}).
Bardeen's Eq.(bd7), from which Eq.(bd52)=Eq.(\ref{bardeen0}) is obtained, is
\begin{equation}
\frac{1}{\tau_{\rm B}}=\frac{k_{\rm m}^{2}}{\pi\hbar (dE/dk)_{\rm m}}
 \int_{0}^{\pi}|M_{\rm kk'}|\rm^{2}(1-\cos\theta)\sin\theta d\theta,
             \label{bd-M1}
\end{equation}
where $|M_{\rm kk'}|$ is the transition matrix for
$\bf k\rightarrow k'\,[\,\cos\theta=(\bf{k\cdot k'})/\rm kk'$ 
and $\bf{k}$ is the vector parallel to
the applied electric field].
In order to see the magnitude of $|M_{\rm kk'}|\rm^{2}$, we first convert 
the factor before the integration in Eq.(\ref{bd-M1}) to $k_{\rm}m/\pi\hbar^{3}=3\pi n/2\hbar E_{\rm F}$. Then
we obtain, assuming that $\tau_{\rm B}=\hbar/k_{\rm B}T$ holds exactly,
\begin{equation}
<\!|M_{\rm kk'}|\rm^{2}\!>_{\theta}\equiv 
\int_{0}^{\pi}|\it M_{\rm kk'}|\rm^{2}(1-\cos\theta)\sin\theta d\theta
=\frac{2\it E_{\rm F}k_{\rm B}T}{\rm3\pi\it n}.    \label{bd-M2}
\end{equation}
If $|M_{\rm kk'}|$ vanishes below $\theta<\pi$, the upper integration limit can be $\theta<\pi$.
See below a different normalization form at the section end. 


Conventionally $M_{kk'}$ is expressed by a product of the structure factor $S_{\rm t}$
and the form factor $m_{kk'}$, namely  $M_{kk'}=S_{\rm t}m_{\rm kk'}$.
Here $|S_{\rm t}|^{2}=|q|^{2}|a_{q}|^{2}=\frac{3}{2}k_{\rm B}T/E_{\rm F}$, where the first equality is from
Rossiter's\cite{rossiter} Eqs.(5.9b) and (A19) \,($|u_{q}|\equiv |a_{q}|$ is the vibration amplitude) and
the second is from $|a|^{2}=k_{\rm B}T/2Mq^{2}C_{\rm s}^{2}$ of Eqs.(bd9-10) and present Eq.(\ref{bohm}).  
To be consistent with Eq.(\ref{bd-M2}), we need to adopt $<|m_{kk'}|^{2}>_{\theta}=(\frac{2}{3}E_{\rm F})^{2}/\pi n$.
Thus we can express $n_{\rm eff}/n$ in terms of $<|M_{\rm kk'}|^{2}>_{\rm \theta}$ or $|S_{\rm t}|^{2}$ as 
      
\begin{equation}
\frac{n_{\rm eff}}{n}\equiv\alpha\frac{k_{\rm B}T}{E_{\rm F}}
    =\alpha \frac{3\pi n<|M_{\rm kk'}|^{2}>_{\theta}}{E_{\rm F}^{2}}\approx 2|S_{\rm t}|^{2}. 
            \label{bd-M3}
\end{equation}
Similarly the amplitudes of ion vibration is $|a_{\rm q}|/(d/2)=0.63(k_{\rm B}T/E_{\rm F})^{1/2}$,
again only as functions of $T$ and \EF
\,[\,
note that $a_{\rm q}/(d/2)=0.04$ at 300K in Cu]. By an appropriate normalization to the wave function,
using azimuth angle $\phi$ and $\int_{0}^{2\pi}d\phi/\int_{0}^{\pi}(1-\cos\theta)\sin\theta d\theta=\pi,$ Eq.(\ref{bd-M2})
changes to 2\EF\kB$T/3$, and $<|m_{kk'}|^{2}>_{\theta}$ changes to $(\frac{2}{3}E_{\rm F})^{2}$;
in the latter in modern treatments, $m_{kk'}\approx -\frac{2}{3}E_{\rm F}$ for
$q/2k_{\rm F}=\sin(\theta/2)=0$, while $m_{kk'}\approx 0$ at $q/2k_{\rm F}\approx 1$ 
as shown e.g. in Fig.(5.30) or Figs.(6.5-6) of Rossiter\cite{rossiter}.
\\

\begin{center}
\bf{8. Fluctuation Dissipation Theorem in Interpreting the collision time, Eq.(1)}
\end{center}

Now, examples of the ratio of $\tau_{0}$ from Eq.(2) to $\tau_{\rm obs}$ determined from Eq.(1) using the experimental
resistivity and $G=1$ are $\tau_{0}/\tau_{\rm obs}$=0.88(Na), 0.98(Rb), 1.05(Cu), and 0.95(Au)
at room temperature ranges, $T_{\rm room}$. Because of this remarkable closeness to unity, it is tempting to assume that there may be an extremely simple
explanation for $\tau_{0}=\hbar/k_{\rm B}T$, which is presented in this section, using monovalent
metals, since multivalent metals can be treated by Eq.(2) with the guessed $G$-values and $Z_{\rm \sigma}=1$.

For this purpose, first we search for the connection of $\tau_{0}$ to the fluctuation dissipation theorem, 
FDT\cite{kubo_et_al, reichl,reif}.
The classical representation of FDT
for the isotropic electrical conductivity $\sigma$ is as Eq.(9.1) in the seminal paper of
Kubo \cite{kubo} 
\begin{equation}
\sigma=\frac{ne^{2}}{k_{B}T}\int_{0}^{\infty}<v_{\rm x}(0)v_{\rm x}(t)>dt. \label {Kubo1}
\end{equation}
Here $v_{\rm x}(t)$ is the fluctuating electron velocities parallel to the given electric field $E_{\rm x}$, and
an obvious relation holds for the autocorrelation
 $<\!\!v_{\rm x}(0)v_{\rm x}(t)\!\!>=<\!\!v_{\rm x}(t_{1})v_{\rm x}(t+t_{1})\!\!>$ for any $t_{1}$ in the assumed stationary stochastic processes.
We assume that the time average of any physical quantity $<A>={\rm lim}_{\small{t\rightarrow \infty}}\small\frac{1}{t}\normalsize\int_{0}^{t}A(t)dt$ is equal to ensemble average $\overline{A}$ (ergodic). 

The thermal fluctuation is of course expected without giving external electric fields,
 and is on the order of $(4f(1-f)/N)^{1/2}$ ,
 where $f(E)$ is the Fermi distribution function and $N$ is the total number of electrons 
in a volume $V$\cite{landau}. 
Thus the fluctuation is occurring around the Fermi energy $E_{\rm F}$ with the width $\approx 3k_{\rm B}T.$ 
Historically the macroscopic counter part originates from the 1928-Nyquist `fluctuation' theorem. 

We split Eq.(\ref{Kubo1}) in two parts by introducing the true electron mass $m$.
\begin{equation}
\sigma=ne^{2}\tau_{\rm cor}/m, \qquad\qquad \\   \label {Kubo2}
\end{equation}
where
\begin{equation} 
\tau_{\rm cor}\equiv\frac{m}{k_{B}T}\int_{0}^{\infty}<v_{\rm x}(0)v_{\rm x}(t)>dt.  \label {Kubo3}
\end{equation}
Thus in order to obtain $\tau_{\rm cor}=\tau_{0}\equiv\hbar/k_{\rm B}T$, 
we `need' (but not yet we `obtain')
\begin{equation}
 H \equiv\frac{\tau_{\rm cor}k_{\rm B}T}{2} 
   =\int_{0}^{\infty}<\frac{1}{2}mv_{\rm x}(0)v_{\rm x}(t)>dt
=\frac{\hbar}{2}. \label{H}
\end{equation}
We then adopt $H=<\frac{1}{2}mv_{\rm x}^{2}\!><\Delta t>$, where $<\Delta t>$
is chosen to reproduce the value of $H$ as the average of the time span of this integration. This is reasonable if we consider that the 
autocorrelation function rapidly decreases as exp$(-t/\tau)$ 
for relaxation processes [or in the (first) Markov processes]. 
We interpret Eq. (\ref{H}) as expressing
\begin{equation} 
H=
<\frac{1}{2}mv_{\rm x}^{2}><\Delta t>\equiv<(\Delta E)_{\rm kin-x}>
<\Delta t>=\frac{\hbar}{2}. \label{H1} 
\end{equation}
Here $<\frac{1}{2}mv_{\rm x}^{2}>\equiv<(\Delta E)_{\rm kin-x}>$
literally means the fluctuating kinetic energy of electrons in the $x$ direction.
If $(\Delta E)_{\rm kin-x}\Delta t=\Delta(p_{\rm x}^{2}/2m)\Delta t=(p_{\rm x}/m)\Delta p_{\rm x}\Delta t=\Delta p_{\rm x} v_{\rm x}\Delta t=\Delta p_{\rm x}\Delta x$ is used,
as in e.g. Sect.10 of Messiah\cite{messiah}, we still obtain,
dropping $<...>$ for simplicity,
\begin{equation} 
H=(\Delta E)_{\rm kin-x}\Delta t=\Delta p_{\rm x}\Delta x
   =\frac{\hbar}{2}.   \label{H2} 
\end{equation}

Conversely, assume that $H=<\!\!\frac{1}{2}mv_{\rm x}^{2}><\Delta t>\!\!$ takes the minimum uncertainty value of
$\hbar/2$ as suggested in Eq. (\ref{H1}) or Eq.(\ref{H2}), then Eq.(\ref{H})
immediately leads to $\tau_{\rm cor}=\hbar/k_{\rm B}T=\tau_{0}$.
It seems rather difficult to escape from this statement. 
 
We remark below some relations among Eqs.(\ref{Kubo1})-(\ref{H2}). 
If we assume Eqs.(\ref{Kubo2}) and (\ref{Kubo3}), we obtain
Eq.(\ref{Kubo1}), which may be regarded as a simplest derivation of the 
fluctuation-dissipation theorem in the classical form if so wished.
[In this logic Eq.(\ref{Kubo2}) with an unspecified $\tau$ can be obtained
from the Boltzmann equation (Appendix), and we regard that this $\tau$ is specified (or defined) by Eq.(\ref{Kubo3}).]
Or else from Eqs.(\ref{Kubo1}) and (\ref{Kubo3}), we obtain 
Eq.(\ref{Kubo2}), the Drude form with $\tau=\tau_{\rm cor}$! 
Further, if we assume
\begin{equation}
<\Delta t>=\tau_{\rm cor}    \label{tau-eq-delta-tau}
\end{equation}
and use $\tau_{\rm cor}=m <v_{\rm x}^2><\Delta t>/{k_{\rm B}T} $ from Eq.(\ref{H})
and Eq.(\ref{H1}), 
we find one dimensional equipartition 
\begin{equation}
 \frac{1}{2}m<v_{\rm x}^{2}>=\frac{1}{2}k_{\rm B}T,\label{equi-partition}
\end{equation}
which is quite reasonable, in view of the fact that the fluctuation
 is occurring around $E_{\rm F}$. In fact derivation of the 
classical Nyquist theorem for macroscopic electric circuits uses this equipartition at $T_{\rm room}$ range.  
In the above $<\!\!\frac{1}{2}mv_{\rm x}^{2}\!\!>$ which is non-negative
 comes actually from 
$(\Delta E)_{\rm rms}\equiv<\!(E_{\rm kin-x}-\frac{1}{3}E_{\rm F}\!)\!^{2}\!>^{1/2}
\,\equiv\,\,<(\Delta E_{\rm kin-x}\!)^{2}>^{1/2}
 =\frac{1}{2}m<v_{\rm x}^{2}>$. Thus we obtain $(\Delta E)_{\rm rms}=k_{\rm B}T/2$. 
For a metal in a heat bath,
since the temperature entering $\tau_{0}=\hbar/k_{\rm B}T$ is by itself a
fluctuating quantity expressed as $<(\Delta T)^{2}>^{1/2}/T=(k_{\rm B}/C_{\rm V-lattice})^{1/2}=1/(3N)^{1/2}$
(Landau and Lifshitz\cite{landau}, Chap.12, Eq.(112.6); $N$ is the total number of ions in a Volume $V$),  
$\Delta t$ may be expressed as $<\tau^{2}>^{1/2}=<(\Delta t)^{2}>^{1/2}\equiv (\Delta t)_{\rm rms}$.

Now, our task is to understand why the empirically supported 
$\frac{1}{2}\int_{0}^{\infty}<mv(0)v(t)>dt\equiv<\frac{1}{2}mv^{2}>\Delta t=\frac{\hbar}{2}$ 
holds; namely Eq.(\ref{H1}) or Eq.(\ref{H2}). 
Our proposal is as follows. Consider one electron heading toward $x$ direction with the
speed of \VF=(2\EF$/m)^{1/2}$, then it effectively collides with an ion
until it passes e.g. $180\times$ lattice-distance in Cu at 300K $(L_{\rm mfp}/d=V_{\rm F}\tau_{0}/n^{-1/3}=\frac{2}{3}\frac{E_{\rm F}}{k_{\rm B}T}:
L_{\rm mfp}=V_{\rm F}\tau_{0}$ is the mean free pass).
Namely electrons are nearly free. The elementary quantum physics teaches that
the wave function of a free electron is expressed as
\begin{equation} 
   \it{\Psi(x)}=\frac{\rm 1}{a^{\rm 1/2}\pi^{\rm 1/4}}\exp\left(-\frac{(x-x_{\rm 0})^{2}}{{\rm 2}a}+\frac{ip_{\rm 0}x}{\hbar}\right).
                             \label{min-wave-funct}
\end{equation}
This, in the words of Leighton\cite{leighton} p.105, exhibits the minimum uncertainty product $\Delta x\Delta p=\hbar/2,$
hence Eq.(\ref{H2}) holds. Eq.(\ref{min-wave-funct}) is regarded to the classical statement that a one-dimensional particle is
located at coordinate $x_{\rm 0}$ and is moving with momentum $p_{\rm 0}$.
Further detail\cite{leighton} is in p.98; $(\Delta x)^{2}=a^{2}/2, (\Delta p)^2=\hbar^{2}/2a^{2},
<x>=x_{0},<p>=p_{0},
\,\phi(p)=\frac{1}{\pi^{1/4}}\left({\frac{a}{\hbar}}\right)^{1/2}
\exp[-(p-p_{0})^{2}a^{2}/2\hbar^{2}]\exp[-i(p-p_{\rm 0})x_{0}/\hbar],$
where $\phi(p)$ is the wave function for momentum.    

The above result is strengthened by using 
the Bloch form of the wave function ${\it \Psi}$ in the time
independent \Schrodinger\, equation $H {\it \Psi}=(-\frac{\hbar^{2}}{2m}\nabla^{2} +U({\bf r})){\it\Psi}=E{\it \Psi}$, where $U({\bf r})=U({\bf r+b})$ 
is the potential energy within one-electron approximation. 
\begin{equation}
 {\it\Psi}=u(\bf{r})\exp (\it i\zeta)=u(\bf{r+b})\exp (\it i\zeta)
\label{wave-funct1}
\end{equation}
Here $|\bf b|$ is the lattice distance of order of $d$ and $\zeta\equiv \bf{k}\cdot\bf{r}=p/\hbar\cdot r$ 
; that is the plane wave, i.e. free electrons, modified by the periodic lattice potential.
The average of $\it{\Psi}$ becomes 
\begin{equation}
  <\it{\Psi}>=<u><\exp(i\zeta)>\approx<u>\exp(-\frac{<\zeta^{\rm 2}>}{\rm 2})
  \label{wave-funct2}
\end{equation}
for a real variable of $\zeta$, where $|\zeta|<1$ and $\exp (-i\zeta)=\exp(i\zeta)\rm$, i.e. even probability (see e.g. Appendix C of Rossiter\cite{rossiter}, 
though a misprint corrected below; see also  Eqs.(2.105-107) of Ziman\cite{ziman} on Debye-Waller factor, which seems valid without
the condition $|\zeta|<1$). 
Here an approximation in Eq.(\ref{wave-funct2}) comes from 
\begin{eqnarray}
<\exp(i\zeta)>=<1+i\zeta-\frac{\zeta^{2}}{2}-i\frac{\zeta^{3}}{6}+\frac{\zeta^{4}}{24}-...>\\
 =1-\frac{<\zeta^{2}>}{2}+\frac{<\zeta^{4}>}{24}-... \quad (\rm{even\, function}) \label{exp-funct1}
\end{eqnarray}
and 
\begin{equation}
\exp(-\frac{<\zeta^{2}>}{2})=1-\frac{<\zeta^{2}>}{2}+\frac{<\zeta^{4}>}{8}-...,
\label{exp-funct2}
\end{equation}
where accuracy is $\sim 3\%$ for $\zeta=0.8$ or $0.5\%$ for $\zeta=0.5$, using $\frac{\zeta^{4}}{8}-\frac{\zeta^{4}}{24}=\frac{\zeta^{4}}{12}$.
  
This means that the wave function takes the minimum wave function say in $x$ direction, where the wavelength $\lambda=2\pi/k$
should be large enough so that $\lambda \gg d$ ($d$=the mean atomic distance) requires 
$\zeta=kx\ll x/d<1$ to satisfy the approximation imposed above.
Thus the allowed extension of the present minimum wave function should be $x<d$, though
the condition of $u({\bf r})=u({\bf r+b})$ in Eq.(\ref{wave-funct1}) allows us that it can be enormously extended.
To recall the derivation in a most elemental way, we adopt 
\begin{equation}
\phi(x)=(a/\pi)^{1/4}e^{-ax^{2}/2},
\end{equation}
where $\int|\phi(x)|^{2}dx=1$ and $a>0$.
We obtain
\begin{equation}
<(x-<x>)^{2}>\equiv<(\Delta x)^{2}>=\int x^{2}|\phi(x)|^{2}dx=\frac{1}{2a}, 
\end{equation}
using $<x>=0$. 
The corresponding wave function for $p$ which is the Fourier transform of $\phi(x)$ is calculated to be 
$\psi(p)=\int_{-\infty}^{\infty}\phi(x)\exp(ixp/\hbar)dx
=e^{-p^{2}/2a\hbar^{2}}/(\pi a\hbar)^{1/4}$. 
We similarly obtain 
\begin{equation}
<(\Delta p)^{2}>=\frac{a\hbar^{2}}{2}.
\end{equation}
Hence 
$<(\Delta x)^{2}>^{1/2}<(\Delta p)^{2}>^{1/2}=\frac{\hbar}{2}$
is derived from Eqs.(35-36). [In the pure classical wave of $\psi(k)=\int\phi(x)\exp(ikx)dx$, we naturally obtain 
$<(\Delta x)^{2}>^{1/2}<(\Delta k)^{2}>^{1/2}=\frac{1}{2}$, without $\hbar$.]
The above derivations of $\tau_{0}=\hbar/k_{\rm B}T$ do not use the ion vibration,
since Eq.(\ref{Kubo1}) is independent of it, namely only the electron motions, though Eq.(\ref{bohm}) acts for this connection. 

The conclusion of this section is that if nearly free electrons are assumed,
$\Delta x\Delta p_{\rm x}=\hbar/2$ is satisfied. This means that 
Eq.(\ref{H2}) or Eq.(\ref{H1}) should be satisfied. In turn, Eq.(\ref{H}) from the fluctuation dissipation theorem should hold.
Thus we obtain Eq.(1), assuming $\tau_{\rm cor}=\tau_{0}$.\\

\begin{center}
\bf{9. Discussion}
\end{center}

We remark on the observed temperature dependence of the electrical resistivity $\rho=1/\sigma.$
If we plot $\rho/T$ as a function of $\Theta$, 
where $\Theta$ is the Debye temperature, then we find that $\rho/T=$ constant applies
only in the following small range: 
\begin{equation}   
T\leq 0.5T_{\rm melt}         \label{T-room1}
\end{equation}
\begin{equation} 
0.5 \leq T/\Theta\leq 2    \label{T-room2}
\end{equation}

\noindent Here $T_{\rm melt}$ is the melting point. 
Surprisingly, $T$ beyond $\sim 0.5T_{\rm melt}$,
suddenly $\rho$ changes to behave as $\rho\propto T^{\sim 2}$ 
so that the upper end of 2 in Eq.(\ref{T-room2}) results,
though dependent upon various 19 metals plotted (see Hirayama\cite{hirayama},Fig.1).
; detailed plot for each metal see Bass\cite{bass-landolt-bornstein}} in 1984-Landolt-B$\ddot{\rm o}$rnstein.
The low side of 0.5 in Eq.({\ref{T-room2}) is of cource
the beginning of the influence of Gr${\rm \ddot{u}}$neisen-Bloch-like form $\rho
\propto T^{m}$, where $m$ is not necessarily 5, but only $m\sim 5$. 
Thus what $T_{\rm room}$ implied for $\sigma\propto T$ in this paper
must be understood to be treating the above range, while $C_{\rm V-lattice}/T$=const
keeps generally up to $T_{\rm melt}$ as in Fig.2 of Hirayama\cite{hirayama}. (One interesting 
point in the latter figures is that a unit of $\hbar\Omega_{\rm ion}/k_{\rm B}$ with $Z=1$ instead of the Debye
temperature can be taken another unit temperature.)
Also theoretical treatments using the word `high temperature', or even $T\gg \Theta$
must keep these in mind. 

Bass et al.(1990)\cite{bass} reviewed that Alkali metals show $\rho\equiv 1/\sigma\propto T^{2}$ at $T\le$1K and 
$\rho\propto T$ in regions of $\Theta (\rm Debye\it)\le T \le T_{\rm melt}$ (melting point), which gradually tends to show the Bloch-Gr$\ddot{\rm u}$neisen relation $\rho\propto T^{5}$ in $T\ge$ 20K.
As for the theoretical absolute values at 295K, they claim that
$|\rho_{\rm calc}/\rho_{\rm obs}-1| \le 1\%$ in K and Na are 
reproduced, while in Li this is 180$\%$  and in Cs and Rb it is 10 to 20$\%$. 
On the other hand, in transition
metals the situation is very much worse, or no calculations as summarized by Rossiter\cite{rossiter}.
There are some computations for transition metals, where e.g. the experimental
phonon state density (not for electron) is used as summarized in Rossiter\cite{rossiter}, which give a reasonable agreement though accuracy is about the same as
our Fig. 1b-upper. Whether $\rho\propto T^{5}$ or $T^{2}$ (often referred to electron-electron scattering regime) is dominant at the very low temperature ranges is of the current interest.
We recall the well-known fact that the elastic
 scattering of electrons with ions at high temperature range
of $T\approx T_{\rm room}$ is related to large deflection angles
 $|\theta|\approx \pi$, while in $T\ll T_{\rm room},
\, |\theta|\ll\pi$ leads to
the limiting Gr$\ddot{\rm u}$neisen-Bloch formula of 
$\sigma\propto T^{-5}$
; thus approximately $\sigma\propto T^{-1}\sin^{4}(\theta/2)\propto
T^{-1}(T/\Theta)^{-4}$, namely $\sigma\propto\tau_{0}(T/\Theta)^{-4}\propto(T/\Theta)^{-5}$,
though detailed examination is beyond the scope of the present paper.\\

\begin{center}
\bf{10. Conclusions}
\end{center}  
In our view, the reason why Cr($1/\sigma_{\rm obs} $
=129n$\Omega\,$m=
12.9$\times10^{-6}\Omega\,$cm at $T_{\rm obs}$=295K ) 
is more resistive than Cu (17n$\Omega\,$m, 4s$^{1}, G$=1), in fact by 7.6, is simply due to large 
$G$(Cr, 3d$^{5}$4s$^{1}$)=6, i.e. a larger number of equally contributing bands, 
since $n_{\rm atom}$(Cr)=8.3 is nearly equal to 
$n_{\rm atom}$(Cu)=8.5 (10$^{28}$m$^{-3}$ unit).

We find in this paper that Eq.(2) for $\sigma$ and $\lambda$ using $\tau_{0}=\hbar/k_{B}T$ of Eq.(1) agrees well with the
experiments for most of elemental metals, inclusive of semimetals, at room temperature ranges. 
Though only for the thermal conductivity, even semiconductors of Si and Ge agree with our formulae.
Here we adopt that $G$ is the sum of outer electron numbers in the 
electron configuration with some modifications
(e.g. 3d$^{10}$4s is replaced by 4s in Cu). Also n=\natom, namely 
$Z=1$, and $m^{*}=m$ are found to be consistent with the $\sigma$-observations.
These results are found to be, in the words of classical equation of motion for a single
electron, $mdV_{x}/dt=-eE_{\rm x}-m\nu_{\rm tot}V_{\rm x-drift}=0$. Here $\nu_{\rm tot}$ is the
total resistive frequency given by $\nu_{\rm tot}=\Sigma \nu_{i}=\frac{1}{\tau_{0}}G$, where every
$\nu_{i}$ for each $i$-band is assumed equal to $1/\tau_{0}$. 

Three versions of the theoretical derivation of $\tau_{0}=\hbar/k_{\rm B}T$ are considered at the room temperature range. (A) we find that the relaxation time from the Bardeen
is $\tau_{\rm B}=\tau_{0}\times(C/E)^{-2} (C=$deformation ion potential), if we accept the observationally found
near-Bohm-Staver relation of $3MC_{\rm tot}^{2}=$\EF\, of Eq.(\ref{bohm}), where $C_{\rm tot}=
(C_{\rm long}^{1/3}+2C_{\rm trans}^{1/3})^{3}$ is the averaged tolal sound speed for each ion of mass $M$.
This takes care of the often claimed assertion that the ion vibration is a single source of the
non-zero resistivity.
Because $\sigma/\sigma_{\rm obs}=1.0\pm 0.2$ or so as in Fig.1(b)-upper  is found for the majority
of metals, we better conclude that in fact $(C/E)^{-2}$ should be within that order, or simply unity.
(B) we assume $\tau_{0}=n_{\rm eff}S$\VF, where $n_{\rm eff}=(3.3kT/E_{\rm F})n$ is from the steeply decreasing part
of the electron Fermi distribution function, and $S$ is the effective collisional area.
Here we find $\pi a_{0}^{2}<S<d^{2}\equiv n^{-2/3}$, so that because we know 
$S\approx d^{2}/2$ from $\tau_{0}$, it is not a bad approximation, but also beacuse of
its simplicity, it may help for further study with regard to (A) or (C). 

As a third derivation of (C), we take a split form of Eq.(\ref{Kubo2}),
$\sigma=ne^{2}\tau_{\rm cor}/m $, 
and Eq.(\ref{Kubo3}) from
the classical form of the fluctuation dissipation theorem of Eq.(\ref{Kubo1}). If we assume that the electrons 
are nearly free as evidenced from the large mean-free-paths, then the wave packet
in the \Schrodinger\, equation takes $\Delta x\Delta p_{\rm x}=\hbar/2$, which is
the minimum uncertainty value, regardless of whether one takes a completely free-atom or a Bloch form. Or similarly $\Delta E_{\rm kin-x}\Delta{t}=\hbar/2$\cite{messiah}.
\begin{equation}
\tau_{\rm cor}\equiv\frac{\int_{0}^{\infty}<v_{\rm x}(0)v_{\rm x}(t)>dt}{k_{\rm B}T/m}
=\frac{\int_{0}^{\infty}\frac{m}{2}<v_{\rm x}(0)v_{\rm x}(t)>dt}{k_{\rm B}T/2} \label{taucor1}\\
\end{equation}
\begin{equation}
=\frac{2}{k_{\rm B}T}<\frac{1}{2}mv_{\rm x}(0)^{2}><\Delta t> \label{taucor2}
\end{equation} 
\begin{equation}
=\frac{2}{k_{\rm B}T}<\Delta E_{\rm kin-x}><\Delta t> 
=\frac{2}{k_{\rm B}T}\frac{\hbar}{2}
=\frac{\hbar}{k_{\rm B}T}  \label{taucor3}
\end{equation}
Here Eq.(\ref{taucor1})$\equiv$ Eq.(\ref{Kubo3}), Eq.(\ref{taucor2}) takes the average of the autocorrelation function,
and in Eq.(\ref{taucor3}), $\Delta E_{\rm kin-x}$ is the fluctuating kinetic energy of electrons.
If we use $\Delta E_{\rm kin-x}\Delta t=\Delta p_{\rm x}\Delta x=\hbar/2$, the same resullt will hold. This method again explicitly uses no ion-vibrations (of course, Eq.(\ref{bohm}) counts).
The method (C) involves no adjustable parameters, matches the observations quite well as Fig.1(b)-upper, and hence is perhaps a most reliable.
 

I thank Profs. Noboru Miura, Toru Suemoto, and Hiroshi Ezawa for discussion. 



\appendix
\label{App}
\section{\bf{Derivation of Conductivity from the Boltzmann Equation}}

This appendix is only to recall the well-known derivation of Eq.(2) for $\sigma$
without specifying $\tau$ and using $G=1$, electron number density $n=n_{\rm atom}$ (Sect.4), and 
$ m=m^{\ast}$ (Sect.5).
Thus we derive $\sigma=j_{x}/E_{x}$
for a single band as  preparation for multi-band metals:
e.g. each one of five `d' or one `s' band in Cr(3d$^{5}$4s${^1})$. 
We rewrite the current density as
\begin{equation}
j_{x}\!=\!-e\int v_{x}fd\bf k_{\rm 1}
\!=-\it e\int\frac{\hbar k_{x}}{m}(f\!-\!f_{\rm 0})d\bf k_{\rm 1}, \label{current-density}
\end{equation}
using $d{\bf k}_{\rm 1}\equiv g_{\rm e}d{\bf k}/(2\pi)^{3}$\,@
($g_{\rm e}=2$ is from electron spin), $\it mv_{x}\!=\!\hbar k_{x}$, and 
$f_{0}$ and $f $ are  unperturbed and perturbed Fermi distribution function,  respectively. 
For convenence, to the last term we added  $-\int_{-\infty}^{\infty}\it \frac{\hbar}{m}k_{x}f_{\rm 0}d\bf k_{\rm 1}=\rm 0$,
because $k_{\rm x}f_{0}$ is an odd function of $k_{\rm x}$, and hence vanishes. 
 
Given the electric field $ E_{x}$ in the $x$ direction, the `steady' Boltzmann equation
from $ \partial f/\partial t-eE_{x}\partial f/\hbar\partial  k_{x}=(\partial f/\partial t)_{\rm coll} $
is
\begin{equation}
  -eE_{x} \frac{\partial f_{0}}{\hbar\partial  k_{x} }=-\-\frac{f-f_{0}}{\tau},  \label{boltz}
\end{equation}
where we adopted the relaxation time  approximation for the collision term $(\partial f/\partial t)_{\rm coll}$.
First we insert  $(f\!-\!f_{0})$ of Eq.(\ref{boltz}) into Eq..(\ref{current-density})
and assume a $ \bf k$-independent $\tau$ as in Eq.(1).
We thus obtain $\sigma=j_{x}/E_{x}$ in the Drude form as
\begin{equation}
\frac{\sigma}{e^{2}\tau/m}=-\int k_{x}\frac{\partial  f_{0}}{\partial k_{x}}d\bf k_{\rm 1}
=\it \int\!f_{\rm 0}\frac{ d\bf k}{\rm 4\pi^{3}}=n.  \label{sigma-3}
\end{equation}
Partial integration over $dk_{x}$ within the primitive cell  leads to 
the second equality of Eq.(\ref{sigma-3}) ; for   $f=f(k_{\rm x},k_{\rm y},k_{\rm z})$, $(\int_{-\infty}^{\infty}k_{\rm x}\frac{\partial f}{\partial k_{\rm x}}dk_{\rm x})dk_{\rm y}dk_{\rm z}
 =-(\int_{-\infty}^{\infty}fdk_{\rm x}) dk_{\rm y}dk_{\rm z}$, and
$k_{\rm x}f\rightarrow 0$ for $k_{\rm x}\rightarrow \pm\infty$ due to more rapidly changing
$f$ than the change of $k_{\rm x}$.
  The final equality of Eq.(\ref{sigma-3}) leads to $n$, where $n=\int fD(E)dE=\int f_{0}d\bf{k}\rm/4\pi^{3}$ ($D(E)$ is the state density).

Because of the factor $\partial  f_{0}/\partial k_{x}$ in Eq.(\ref{sigma-3}),
 not only  a band, where the state density does not extend beyond the Fermi energy,
does not contribute to $\sigma $ such as 3d$^{10} $ of Cu (see e.g. Ibach and L$\ddot{\rm u}$th\cite{ibach}, Fig.7.12, left),
but also only electrons having the  energy of $ E_{\rm F}  \pm 2k_{\rm B}T\approx  E_{\rm F} $ are 
contributing to $\sigma$  ($E_{\rm F}$=Fermi energy=2-7eV for $Z$=1). 
For the same reason, even if $\tau$ is dependent upon $k$, 
this must come from $\tau=\tau(k\approx k_{\rm F})$.
  
Since we used coordidates of $f=f(k_{\rm x},k_{\rm y},k_{\rm z})$,
the result, namely $n$, is the same as given in Eq.(\ref{sigma-3}) even for
the non-free electron model such as non-hyperbolic bands. In the latter the surface area is left unknown, though useful
for some other problems.
Note that the departure from $d{\bf k}=4\pi k^{2}dk=D_{\rm free}(E_{\rm F})dE$
is by a multiple factor of $D(E_{\rm F})/D_{\rm free}$=1.2-1.6 from the 
tabulation of $D(E_{\rm F})$ for each metal in Moruzzi et al \cite{moruzzi}, and that
the model appears to change the  value of $ E_{\rm F}$. 
 

We adopted
$(\partial f/\partial t)_{\rm coll}=-(f-f_{0})/\tau$ as the collision term. 
This can be understood by $f=f_{0}-(\partial f/\partial t)_{\rm coll}\tau$ of the Taylor
expansion. Therefore, though it is conventionally called `relaxation time 
approximation', in reality, it can be anything that satisfies
the linear Taylor expansion, such as the correlation time $\tau_{\rm cor}$ as shown in Sect.9. The correlation
time may well be more appropriate since we assumed $\partial f/\partial t=0$
in Eq.(A2). Besides,
the linear Taylor expansion is readily shown accurate enough, 
since the required electric field for non-linearity is quite large [ e.g.
Grosso and Parravicini\cite{grosso} below their Eq.(11.30b)].

The rough derivation above may be sufficient for
the discussion in the text.

\end{document}